\documentclass[aps,prb,twocolumn,amsmath,amssymb,superscriptaddress,longbibliography]{revtex4-1}
\usepackage[dvips]{graphicx}
\usepackage{graphicx}
\usepackage[utf8]{inputenc}
\usepackage{dcolumn}
\usepackage{physics}
\usepackage{soul}
\usepackage{bm}
\usepackage{graphics}
\usepackage{dirtytalk}
\usepackage{epsfig,color}
\usepackage{xcolor,cancel}
\usepackage[breaklinks,colorlinks,bookmarks=false,citecolor=blue,linkcolor=magenta,urlcolor=blue]{hyperref}
\graphicspath{{figure/}} 

\begin{document}
\title[J. Vahedi]{Thermovoltage and heat dissipation in a triangle quantum dot junction }

\author{ Ahmad Ahmadi Fouladi}
\email{a.ahmadifouladi@iausari.ac}
\affiliation{Department of Physics, Sari Branch, Islamic Azad University, Sari 48164-194, Iran.}

\author{Javad Vahedi}
\email{j.vahedi@jacobs-university.de}
\affiliation{Department of Physics and Earth Sciences, Jacobs University Bremen, Bremen 28759, Germany.}
\affiliation{Department of Physics, Sari Branch, Islamic Azad University, Sari 48164-194, Iran.}

\date{\today}
\begin{abstract}
We numerically investigate the thermoelectric properties of a triangle quantum dot connected to metallic electrodes using the  non-equilibrium Green's function method in the Anderson model. Exploiting the equation of motion method in the Coulomb-blockade regime, the thermovoltage, thermocurrent and heat dissipation are calculated. Results show that the thermovoltage and thermocurrent have nonlinear behavior, and the magnitude and sign of them can be controlled with site energy and coupling strength of quantum dots. Moreover, we find that the heat current is nonlinear and asymmetric respect to the sign of bias voltage for all of the site energies of quantum dots. Analyses show that the heat current can be positive or negative for all of the site energies and becomes  zero for the nonzero voltages. These results can be useful to determine the performance of the nanoscale electronic devices to control the heat dissipations.
\end{abstract}

\maketitle
\section{Introduction}\label{sec1}
The thermoelectric properties of quantum dots(QDs) have attracted th attention of both theoretical and experimental researchers in recent years~\cite{Sothmann2014,Urban2015, Erdman2017, Talbo2017, De2018, Menichetti2018, Tang2018, 
Ludovico2018, Svilans2018, Josefsson2018, Josefsson2019, Chi2020, Ribetto2021, Jong2021,
Banerjee2021, Dorsch2021, Zimbovskaya2022}. Exploiting nanostructures such as QDs and organic molecules placed between metalic electrodes show high-efficiency energy conversion devices in the nonlinear regime ~\cite{Sierra2014,Zimbovskaya2015,Sierra2015,Zimbovskaya2016,Sanchez2016,Svilans2016,Orellana2018,Sartipi2018,Vahedi2018,Peltonen2019,Taniguchi2020}. Accordingly, studying the thermoelectric properties of nanoscale systems provide a deeper insight into the nature and characteristics of the electron and heat transfer process. 
\par
Heat-to-electric converters operate through the Seebeck effect~\cite{Goldsmid2010}. The Seebeck effect arises when either electric and thermal forces act simultaneously on electron transmission through the desired system. When a temperature difference occurs along the system in the absence of bias voltage application, a thermovoltage $V_{\rm th}$ arises and in this case the electric current is zero. As a result, the formation of $V_{\rm th}$ implies the conversion of energy. In 1993, Staring et al. reported an interesting behavior of $V_{\rm th}$ in a QD in the presence of Coulomb-blockade~\cite{Staring1993}. Their observations showed that  $V_{\rm th}$ increases with the application of temperature bias, which is in agreement with the Seebeck effect. Moreover, at higher temperature biases, $V_{\rm th}$ decreases and then becomes zero for a non-zero temperature difference, and eventually its sign changes. Svensson, et al. also studied the nonlinear properties of temperature voltage in nanowires and obtained a similar result~\cite{Svensson2013}. 
\par
The cause of this nonlinear behavior is attributed to the renormalization of energy levels by temperature; Because the accumulated charge depends on the applied temperature gradient. Unlike the current caused by the application of voltage, which has a definite sign for voltages greater than zero, by applying a temperature gradient, the heat current can be positive or negative, depending on the sign of the heat power. For electron-like charge carriers, it is a positive and for hole-like carriers, it is a negative ~\cite{Sierra2014}. However, in the linear response regime, the heat power is constant and the above reason is not sufficient to change the heat flow sign. Therefore, a strong and negative differential heat conduction is needed to change the flow from positive to negative values~\cite{Sierra2014}. It is interesting that by increasing the temperature of the electrodes, the heat flow through the QD may be zero. Notably, this is a completely nonlinear thermoelectric effect and bears no resemblance to voltage or linear thermoelectric regime.
\par
\begin{figure}[htb]
    \includegraphics[width=0.48\textwidth]{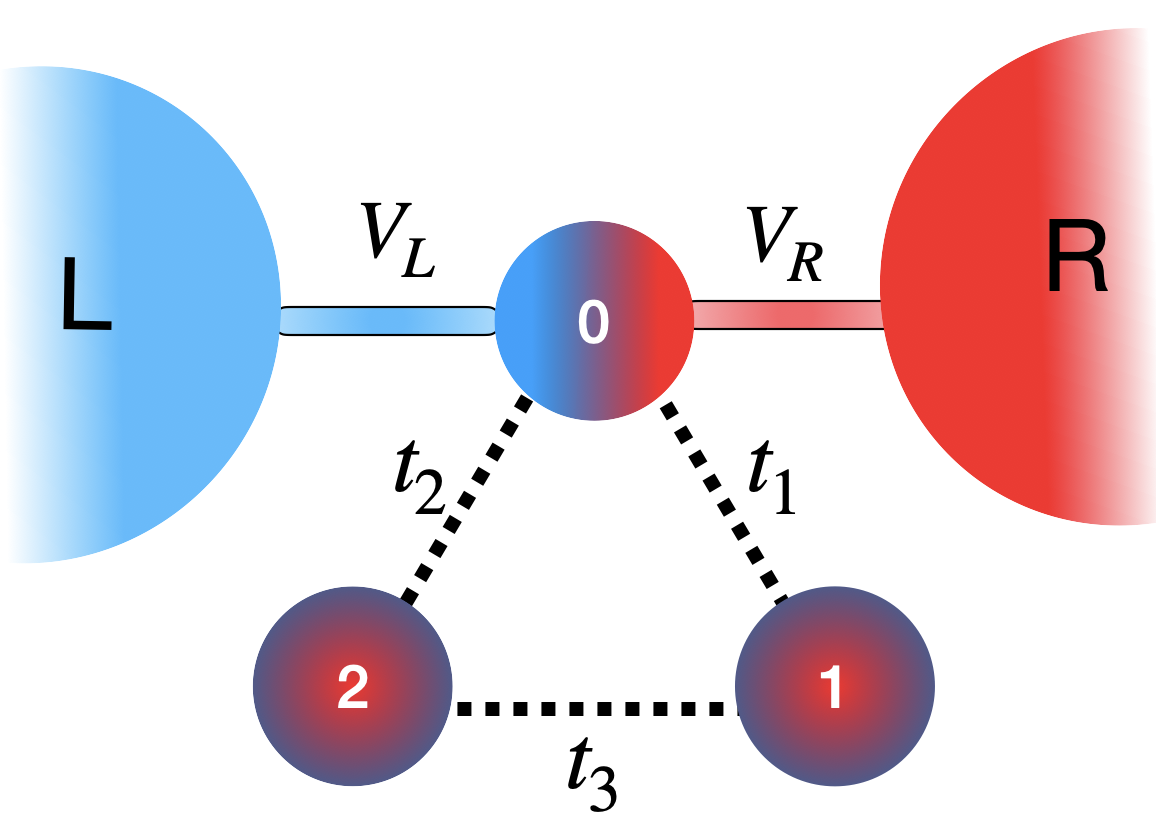}
    \caption{(color online ) Cartoon showing the model with a triangular quantum dot  attached to semi-infinite metallic leads to which both a voltage $\Delta V$ difference and a temperature difference $\Delta T$ are applied.}
    \label{fig1}
\end{figure}

Considering the effect of Coulomb-blockade in low temperature physics is important for nanometer systems~\cite {Wierzbicki,Zianni}. The effect of Coulomb interaction occurs when the energy of the electron-electron interaction becomes appreciable compare to the other energies scale of the model, such as the energies levels broadening and the temperature. Because electrons carry energy in the tunneling process, Coulomb-blockade can have a significant effect on heat conduction~\cite{Zimbovskaya2,Sierra2}. In this paper, combing the non-equilibrium Green function and the equation of motion methods, we study thermovoltage, thermocurrent, and heat dissipation in a triangle QD attached to metallic electrodes, see Fig.~\ref{fig1} for a sketch,  in the presence of electron-electron interaction.
\par
The rest of the paper is organized as follows. In Sec. \ref{sec2}, we first introduce the model and then give a detailed analytical derivation of transport properties with the combination of non-equilibrium Green's function and equations of motions. In Sec. \ref{sec3}, we present numerical results and their interpretations. Conclusion are summarized in Sec.\ref{sec4}.
\section{MODEL and FORMALISM}\label{sec2}
Here we study three QDs setup in as a triangle structure attached to metallic leads. The leads are connected to the ${\rm QD}_0$, as shown in Fig.~\ref{fig1}. Moreover, we consider a bias voltage and temperature gradient between the two leads.  The total Hamiltonian describing the device is given by
\begin{equation}
\mathcal{H} = \mathcal{H}_{L}+\mathcal{H}_{L-dot}+\mathcal{H}_{dot}+\mathcal{H}_{R}+\mathcal{H}_{R-dot},
\label{eq1}
\end{equation}
where
$\mathcal{H}_{L/R}=\sum_{k \sigma}\epsilon_{k\sigma}c^{\dagger}_{k\sigma}c_{
k\sigma}$ corresponds to the left (right) metallic lead. $c^{\dagger}_{k\sigma}$ ($c_{k\sigma}$) creates (annihilates) an electron at state $(k,\sigma)$ in the lead $\alpha=L/R$. The term $\mathcal{H}_{\alpha-dot}= \sum_{k_\alpha\sigma} \big(\mathcal{V}_{k_\alpha\sigma}c^{\dagger}_{k_\alpha\sigma}d_{0\sigma}+h.c\big)$ describes the coupling between the ${\rm QD}_{0}$ and the leads, where $\mathcal{V}_{k_\alpha\sigma}$ denotes the hopping amplitude between ${\rm QD}_{0}$ of the central region and state $k$ of lead $\alpha=L/R$. The term $\mathcal{H}_{dot}$
describes the central region made of three QD in a triangular shape which we model as follows
\begin{eqnarray}
\mathcal{H}_{dot}&=&\sum_{j=0,\sigma}^2\epsilon_{j\sigma}d^{\dagger}_{j\sigma} \, d_{j\sigma}+t_1\sum_\sigma\big(d^{\dagger}_{0\sigma} \, d_{1\sigma}+h.c\big)\nonumber\\
&&~+t_2\sum_\sigma\big(d^{\dagger}_{0\sigma} \, d_{2\sigma}+h.c\big)+t_3\sum_\sigma\big(d^{\dagger}_{1\sigma} \, d_{3\sigma}+h.c\big)\nonumber\\
&&~+U n_{o\uparrow}n_{0\downarrow}\, ,
\label{eq2}
\end{eqnarray}
where $d^{\dagger}_{j\sigma}$ ($d_{j\sigma}$) are fermion creation (annihilation) operators at ${\rm QD}_{j}$, $U$ is  the Coulomb-interaction and the number operator is given by  $n_{0\sigma}=d^{\dagger}_{0\sigma}d_{0\sigma}$ with spin $\sigma$ at ${\rm QD}_{0}$. 
\par
Electron current can be measured through the time evolution of the expectation value  of electron occupation in one of the leads $I_\alpha=-e\frac{d}{dt}\left<n_\alpha\right>$, where $n_\alpha=\sum_{k_\alpha \sigma}d^{\dagger}_{k_\alpha\sigma}d_{k_\alpha
\sigma}$. Since the total density commutes with the Hamiltonian Eq.~(\ref{eq1}), the current conservation requires that in steady-state to have $\sum_\alpha I_\alpha=0$. Hence the current flowing through the system is considered  $I\equiv I_L=-I_R$ . Utilizing the Keldysh formalism~\cite{Meir1992} current through the interacting quantum dot reads as follows
\begin{equation}
I=-\frac{e}{\pi\hbar}\int dE\sum_\sigma\frac{\Gamma_L\Gamma_R}{\Gamma}{\rm \bf{Im}}~G^r_{\sigma,\sigma}(E)\big[f_L(E)-f_R(E)\big]\, ,
\label{eq3}
\end{equation}
where $G^r_{\sigma,\sigma}(E)$ is the model retarded Green function in the presence of both coupling to the continuum states and electron-electron interactions. $\Gamma_\alpha(E)=2\pi\rho_\alpha(E)|\mathcal{V}_{\alpha\sigma}(E)|$ represents the level broadening due to coupling to the leads ( with total line width $\Gamma=\sum_\alpha\Gamma_\alpha$, and the density of states $\rho_\alpha=\sum_k\delta(E-\epsilon_{\alpha k})$ in the $\alpha$ lead). We consider wide band limit in this paper, so $\Gamma_\alpha(E)\equiv\Gamma_\alpha$. We also consider the density of states and the probability of tunneling independent of spin (non-magnetic electrodes).  $f_\alpha(E)=\big[1+\exp(E-\mu_\alpha)/(k_BT_\alpha)\big]^{-1}$ is the Fermi-Dirac function for lead $\alpha$ with electrochemical potential $\mu_\alpha=E_F+eV_\alpha$ and temperature $T_\alpha=T+\theta_\alpha$ ( where $E_F$ is the Fermi energy and $T$ is the background temperature).
\par
As stated before, we adapt the equation-of-motion technique and restrict it to the Coulomb-blockade regime where two energies scale of the model, namely $k_BT$ and $\Gamma$ are much smaller than the $U$. Within such a restriction, ignoring the complex cotunneling process and working in temperature $T>T_K$ larger than the Kondo temperature, the equation-of-motion approach generates a good description of the transport properties of strongly interacting quantum dots.
\par
Retarded green function $G^r(\tau,0)$ of two fermionic operators $A$ and $B$, and the corresponding equation of motion in energy domain given as 
\begin{equation}
G^r_{A,B}(\tau,0)=\ll A,B\gg_{\tau}^r=-i\theta(\tau)\left<\left\{A(\tau),B(0)\right\}\right>\, ,
\label{eq4}
\end{equation}
\begin{equation}
\left(\epsilon+i0^+\right)\ll A,B\gg_{\epsilon}^r+\ll[H,A],B\gg_{\epsilon}^r=\left<\left\{A,B\right\}\right>\nonumber
\label{eq4b}
\end{equation}
where $H$ is the total Hamiltonian Eq.~(\ref{eq1}), and $0^+$ is  an infinitesimal real number. For the sake of simplicity, in the following we omit index $r$ and $0^+$. Then for ${\rm QD}_0$ we have 
\begin{equation}
\epsilon\ll d_{0\sigma},d_{0\sigma}^\dagger\gg=\left<\left\{d_{0\sigma},d_{0\sigma}^\dagger\right\}\right>-\ll  \big[H,d_{0\sigma}\big],d_{0\sigma}^\dagger  \gg
\label{eq5}
\end{equation}
by inserting Eq.~(\ref{eq1}) into the above equation, we get
\begin{eqnarray}
&&(\epsilon-\epsilon_{0\sigma})\ll d_{0\sigma},d_{0\sigma}^\dagger\gg=1+U\ll d_{0\sigma}n_{0,\bar{\sigma}},d_{0\sigma}^\dagger\gg\nonumber\\
&&+\sum_{\alpha_k\sigma} \mathcal{V}_{k_\alpha\sigma}^*\ll c_{k_\alpha\sigma},d_{0\sigma}^\dagger\gg+t_1\ll d_{1\sigma},d_{0\sigma}^\dagger\gg\nonumber\\
&&+t_2\ll d_{2\sigma},d_{0\sigma}^\dagger\gg
\label{eq6}
\end{eqnarray}
where
\begin{equation}
\ll c_{k_\alpha\sigma},d_{0\sigma}^\dagger\gg=\frac{\mathcal{V}_{k_\alpha\sigma}}{
\epsilon-\epsilon_{k_\alpha}}\ll d_{0\sigma},d_{0\sigma}^\dagger\gg
\label{eq7}
\end{equation}
and to find $\ll d_{1(2)\sigma},d_{0\sigma}^\dagger\gg$, we only need to use Eq.~(\ref{eq4}), then we have
\begin{eqnarray}
\ll d_{1\sigma},d_{0\sigma}^\dagger\gg&=&\frac{t_1^*}{\epsilon-\epsilon_{1\sigma}}\ll d_{0\sigma},d_{0\sigma}^\dagger\gg\\
&+&\frac{t_3t_2^*}{(\epsilon-\epsilon_{1\sigma})(\epsilon-\epsilon_{2\sigma})}\ll d_{0\sigma},d_{0\sigma}^\dagger\gg\nonumber\\
&+&\frac{|t_3|^2}{(\epsilon-\epsilon_{1\sigma})(\epsilon-\epsilon_{2\sigma})}\ll d_{1\sigma},d_{0\sigma}^\dagger\gg\nonumber\, ,
\label{eq8}
\end{eqnarray}
\begin{eqnarray}
\ll d_{2\sigma},d_{0\sigma}^\dagger\gg&=&\frac{t_2^*}{\epsilon-\epsilon_{1(2)\sigma}}\ll d_{0\sigma},d_{0\sigma}^\dagger\gg\\
&+&\frac{t_3^*t_1^*}{(\epsilon-\epsilon_{1\sigma})(\epsilon-\epsilon_{2\sigma})}\ll d_{0\sigma},d_{0\sigma}^\dagger\gg\nonumber\\
&+&\frac{|t_3|^2}{(\epsilon-\epsilon_{1\sigma})(\epsilon-\epsilon_{2\sigma})}\ll d_{2\sigma},d_{0\sigma}^\dagger\gg\nonumber\, ,
\label{eq9}
\end{eqnarray}
with more simplification and by defining  two new parameters 
\begin{eqnarray}
\mathbf{\Sigma}_1&=&\frac{t^*_1(\epsilon-\epsilon_{2\sigma})+t_3t_2^*}{(\epsilon-\epsilon_{1\sigma})(\epsilon-\epsilon_{2\sigma})-|t_3|^2}\nonumber\\
\mathbf{\Sigma}_2&=&\frac{t^*_1(\epsilon-\epsilon_{2\sigma})+t_3t_2^*}{(\epsilon-\epsilon_{1\sigma})(\epsilon-\epsilon_{2\sigma})-|t_3|^2}\nonumber\, ,
\end{eqnarray}
we find a compact form $$\ll d_{1(2)\sigma},d_{0\sigma}^\dagger\gg=\mathbf{\Sigma}_{1(2)}\ll d_{0\sigma},d_{0\sigma}^\dagger\gg,$$ where by substtituation it and Eq.~(\ref{eq7}) into Eq.~(\ref{eq6}) we arrive
\begin{eqnarray}
&&(\epsilon-\epsilon_{0\sigma})\ll d_{0\sigma},d_{0\sigma}^\dagger\gg=1+U\ll d_{0\sigma}n_{0,\bar{\sigma}},d_{0\sigma}^\dagger\gg\nonumber\\
&&+\left(\sum_{\alpha_k\sigma} \frac{|\mathcal{V}_{k_\alpha\sigma}|^2}{\epsilon-\epsilon_{\alpha_k\sigma}}+\sum_{i=1,2}t_i\mathbf{\Sigma}_i\right)\ll d_{0\sigma},d_{0\sigma}^\dagger\gg
\label{eq10}
\end{eqnarray}
with recasting and operating the wide-band approximation $\mathbf{\Sigma}_0=\sum_{\alpha_k\sigma} \frac{|\mathcal{V}_{k_\alpha\sigma}|^2}{\epsilon-\epsilon_{\alpha_k}}\simeq\frac{\Gamma}{\pi}\ln{\left|\frac{D+\epsilon}{D-\epsilon}\right|}-i\frac{\Gamma}{2}$ the Eq.~(\ref{eq10}) can be read as
\begin{eqnarray}
\left(\epsilon-\epsilon_{0\sigma}-\mathbf{\Sigma}_0-\sum_{i=1,2}t_i\mathbf{\Sigma}_i\right)\ll d_{0\sigma},d_{0\sigma}^\dagger\gg=\nonumber\\
1+U\ll d_{0\sigma}n_{0,\bar{\sigma}},d_{0\sigma}^\dagger\gg
\label{eq11}
\end{eqnarray}
In the above fomula still we need $\ll d_{0\sigma}n_{0,\bar{\sigma}},d_{0\sigma}^\dagger\gg$.  Using Eq.~(\ref{eq4}) and ignoring higher-order correlation terms, we find
\begin{eqnarray}
\left(\epsilon-\epsilon_{0\sigma}-U \right)&&\ll d_{0\sigma}n_{0,\bar{\sigma}},d_{0\sigma}^\dagger\gg\\
-&&\sum_{k_\alpha\sigma}\mathcal{V}_{k_\alpha\sigma}\ll c_{k_\alpha\sigma}n_{0,\bar{\sigma}},d_{0\sigma}^\dagger\gg\nonumber\\
-&&\sum_{i=1,2}t_i\ll d_{i\sigma}n_{0,\bar{\sigma}},d_{0\sigma}^\dagger\gg\approx\left<n_{0\bar{\sigma}}\right>\nonumber\, ,
\label{eq12}
\end{eqnarray}
where
\begin{eqnarray}
\ll c_{k_\alpha\sigma}n_{0,\bar{\sigma}},d_{0\sigma}^\dagger\gg&=&\sum_{k_\alpha\sigma}\frac{\mathcal{V}_{k_\alpha\sigma}}{\epsilon-\epsilon_{k_\alpha\sigma}}\ll d_{0\sigma}n_{0,\bar{\sigma}},d_{0\sigma}^\dagger\gg\nonumber\\
\ll d_{i\sigma}n_{0,\bar{\sigma}},d_{0\sigma}^\dagger\gg&=&\mathbf{\Sigma}_i\ll d_{0\sigma}n_{0,\bar{\sigma}},d_{0\sigma}^\dagger\gg\, ,
\label{eq13}
\end{eqnarray}

inserting  Eq.~(\ref{eq13}) into Eq.~(\ref{eq12}), we arrive 

\begin{equation}
\ll d_{0\sigma}n_{0,\bar{\sigma}},d_{0\sigma}^\dagger\gg=\frac{\left<n_{0\bar{\sigma}}\right>}{\epsilon-\epsilon_{0\sigma}-\mathbf{\Sigma}_0-\sum_{i=1,2}t_i\mathbf{\Sigma}_i-U}
\label{eq14}
\end{equation}
wrapping all, we end a closed-form Green's function for ${\rm QD}_0$ as 
\begin{eqnarray}
G^r_{\sigma,\sigma}(E)&=&\ll d_{0\sigma},d_{0\sigma}^\dagger\gg\nonumber\\
&=&\frac{1-\left<n_{0\bar{\sigma}}\right>}{\epsilon-\epsilon_{0\sigma}-\mathbf{\Sigma}_0-\sum_{i=1,2}\mathbf{\Sigma}_i'}\nonumber\\
&=&\frac{\left<n_{0\bar{\sigma}}\right>}{\epsilon-\epsilon_{0\sigma}-\mathbf{\Sigma}_0-\sum_{i=1,2}\mathbf{\Sigma}_i'-U}
\label{eq15}
\end{eqnarray}
where $\mathbf{\Sigma}_i'=t_i\mathbf{\Sigma}_i$, $(i=1,2)$. This shows $G^r_{\sigma,\sigma}(E)$ depends on the dot occupation for reversed spin $\bar{\sigma}$, $\left<n_\sigma\right>=\frac{1}{2\pi i}\int dEG^<_{\sigma,\sigma}(E)$. Using the Keldysh equation for lesser green function $G^<_{\sigma,\sigma}(E)=i\left[\Gamma_Lf_L(E)+\Gamma_Rf_R(E)\right]\left|G^r_{\sigma,\sigma}(E)\right|^1$, and adapting the self-consistent approach we close the system of equations. Analogous to charge current, the heat current is derived from  
\begin{eqnarray}
J_\alpha&=&\frac{d\left< \sum_{k_\alpha\sigma}d^{\dagger}_{k_\alpha\sigma}d_{k_\alpha\sigma}\right>}{dt}\\
&=&\sum_\sigma\frac{\Gamma_L\Gamma_R}{\pi\hbar\Gamma}\int dE(\mu_\alpha-E){\rm \bf{Im}}~G^r_{\sigma,\sigma}(E)\big[f_L(E)-f_R(E)\big]\nonumber
\label{eq16}
\end{eqnarray}
which satisfies the Joule law $J_R+J_L=-IV$.
%
\section{RESULTS AND DISCUSSION}\label{sec3}
Now, we present the results obtained based on the formulation of the previous section. We select the level broadening of ${\rm QD}_0$ due to coupling to the leads $\Gamma_L=\Gamma_R=\Gamma_0$ as a unit of energy. Interested in the Coulomb-blockade regime we appoint background temperature energy and the Coulomb interaction as $k_BT=0.1\Gamma_0$ and $U=10\Gamma_0$, respectively. We also consider all hopping integrals equal as $t_i=\Gamma_0$, ($i=1,2,3$).  
 \begin{figure}[htb!]
    \includegraphics[width=0.5\textwidth]{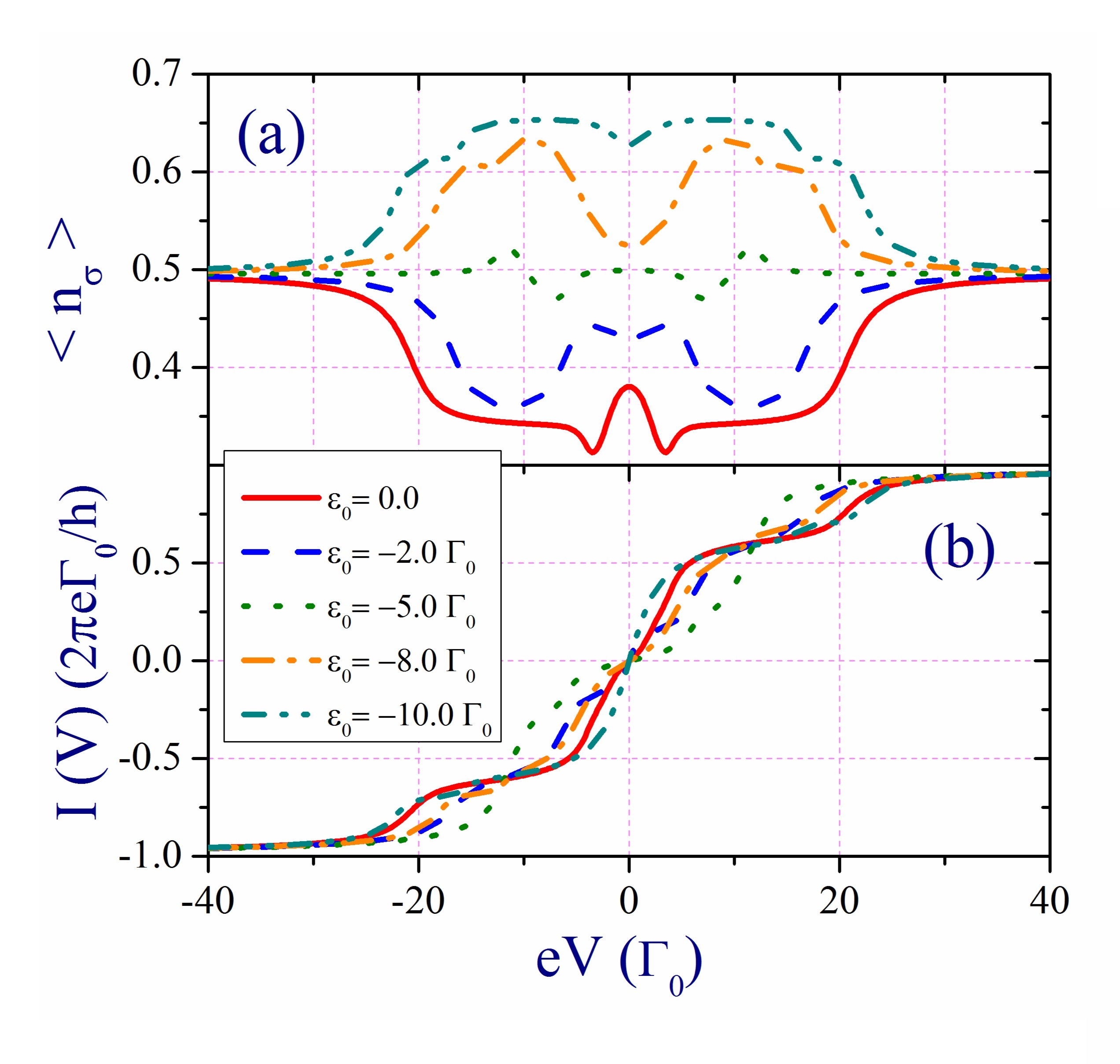}
    \caption{(color online ) (Color online) Panels (a) and (b) show average of occupation number and electrical current  of ${\rm QD}_0$  as function of bias voltage, respectively. Lines correspond to different dot level. Note, we consider a pure voltage-driven process by setting $\theta=0$.}
    \label{fig2}
\end{figure}
\par
Figs.~\ref{fig2}(a) and (b) display the average occupation number and electric current of the ${\rm QD}_0$, respectively, in terms of voltage bias for different ${\rm QD}_0$'s energy values in the absence of temperature gradient ($\theta=0$). The bias voltage applied in a symmetric way to the junction ($\mu_L=-\mu_R=eV/2$), and impose $E_F=0$ as the reference energy point. As it is clear from Fig.~\ref{fig2}(b), for zero bias voltage, when the dot energy level resonates with Fermi energy, the setup acts as an ohmic junction. With increasing voltage, the current shows a step behavior which is a characteristic of the quantum junction and is in accordance with the phenomenological models related to the Coulomb-blockade~\cite{Meir1991,Beenakker1991}. The increase in current occurs when the dot energies of the system are in line with the chemical potential of the electrodes. Furthermore, Fig.~\ref{fig2}(a) displays the level occupation depends strongly on dot level and bias voltage. In a single quantum dot junction, at the particle-hole symmetry point $\epsilon_0=-U/2$, the occupation is voltage independent~\cite{Sierra2014}. It happens since the exchanging $d\rightarrow d^\dagger$ leaves the Hamiltonian of model unaffected. In the triangle quantum dot model defined in Eq.~(\ref{eq1}), particle-hole symmetry is spoiled, as it can be seen from result of $\epsilon_0=-5\Gamma_0$, and shown in a dotted-line on Fig.~\ref{fig2}(a).
\par
\begin{figure}[htb!]
    \includegraphics[width=0.5\textwidth]{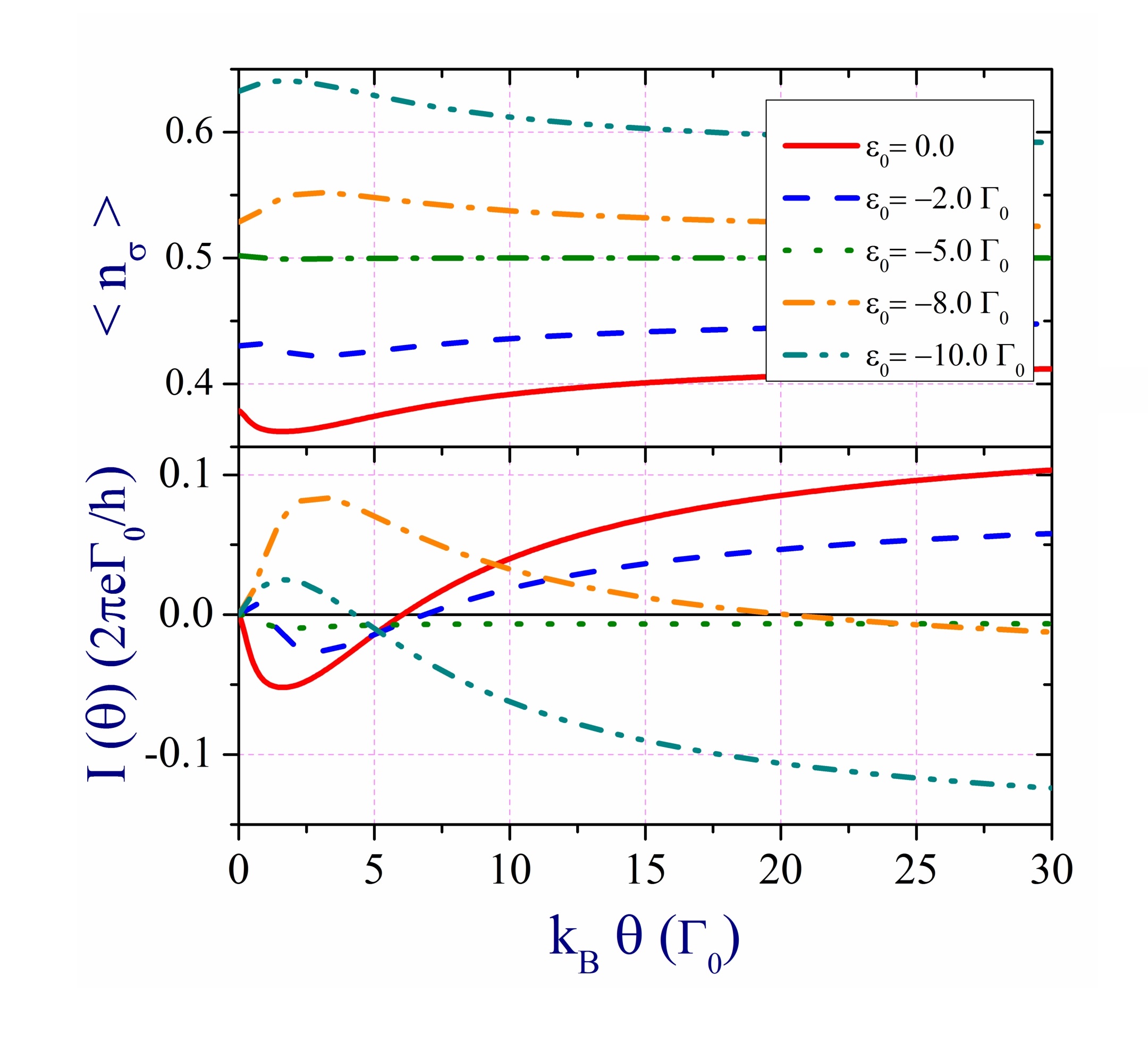}
    \caption{(color online ) (Colore online) Pannels (a) and (b) show average of occupation number and electrical current  of ${\rm QD}_0$  as function of temperature gradiant, respectively. Lines correspond to different dot level. Here, we consider a pure temperature-driven process by setting $V=0$.}
    \label{fig3}
\end{figure}

Fig.~\ref{fig3} displays  the average occupation number and heat current of ${\rm QD}_0$  in terms of the temperature gradient between the two right and left electrodes for different dot energy levels with zero applied bias . Note that in order to have temperature gradient $\theta=\theta_L-\theta_R$, without loss of generality, we set $\theta_L=\theta$ and $\theta_R=0$. In the presence of the temperature gradient, the energy levels renormalize themeself. In this case, there is a displacement in the location of the energy levels of the junction, which can cause a fundamental change in the intensity of electron transport through the junction. Electron flow occurs when the renormalized energy level of the system is close to the chemical potential of the electrodes. The intensity of this flow determines the magnitude of the thermal current.
\par
\begin{figure}[htb!]
     \includegraphics[width=0.5\textwidth]{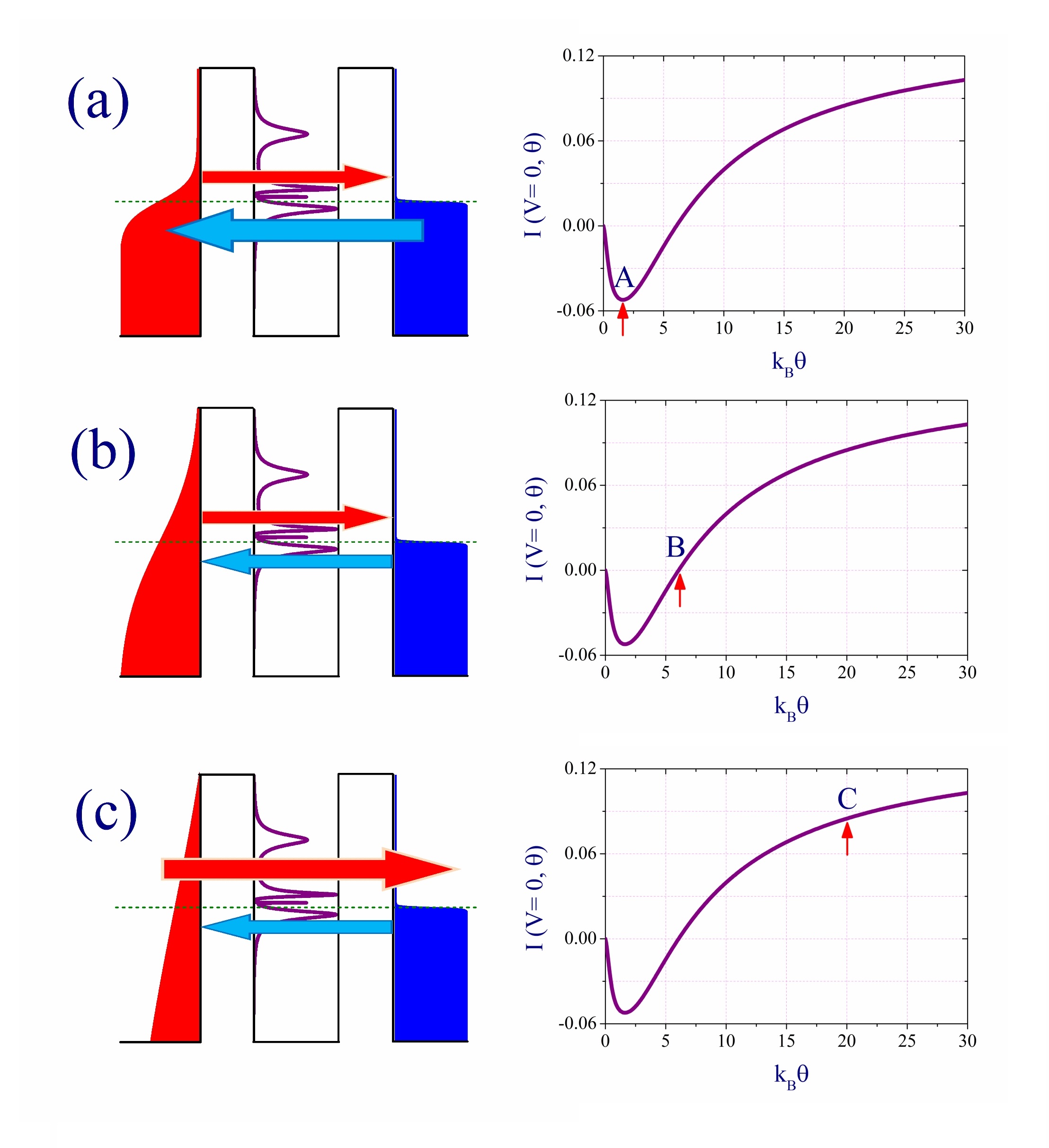}
    \caption{(Color online ) Right panels show a heat current as a function of temperature gradient $\theta$ at dot level $\epsilon=0$. Left panels draw a schematic cartoon of the energy diagram of the transport process through the junction, for three points shown on the heat current plots. }
    \label{fig4}
\end{figure}

For negative (positive) current, the charge carriers are holes (electrons). The results show that the direction of the charge carriers depends on the energy of the ${\rm QD}_0$. For the energy $\epsilon_0\neq-U/2$, the heat current are not zero because  the electronic density of states around the Fermi energy is asymmetric. To have a better insight of the charge carrier transport process, in Fig.~\ref{fig4} we illustrate the heat currents in terms of the temperature difference between the right and left electrodes for dot energy $\epsilon_0=0$. Three left panels (a), (b), and (c) depict a schematic cartoon of the thermally driven transport process through the junction corresponding to points indicated by A, B, and C in the right panels, respectively. At $\theta=0$ heat current is zero, and with increasing $\theta$, the Fermi distribution loses its step function form. The direction of charge flow depends on the dot energy level and the density of state around the Fermi energy(see the horizontal dashed line on the Fig.~\ref{fig4}).  
\par
At point {\bf A}, due to the ordering of energy levels and the density of energy states, it is evident that the contribution of holes in the charge flow is greater than the contribution of electrons, and therefore the heat current is negative. With increasing $\theta$, Fermi distribution extends its tail to higher energies and then make it possible to have more electrons get excited through the junction. By raising the temperature difference more, total heat current again becomes zero, which indicates the balance between the flow of electrons and holes in both directions (see point {\bf B} on the Fig.~\ref{fig4}). Since the number of energy levels available at $E>0$ and dot level $\epsilon_0$ are bigger, increasing $\theta$ leads to higher electrons flow contribution to the total heat current and make it positive, see see point {\bf C} on the Fig.~\ref{fig4}. 
\par
\begin{figure}[htb!]
     \includegraphics[width=0.5\textwidth]{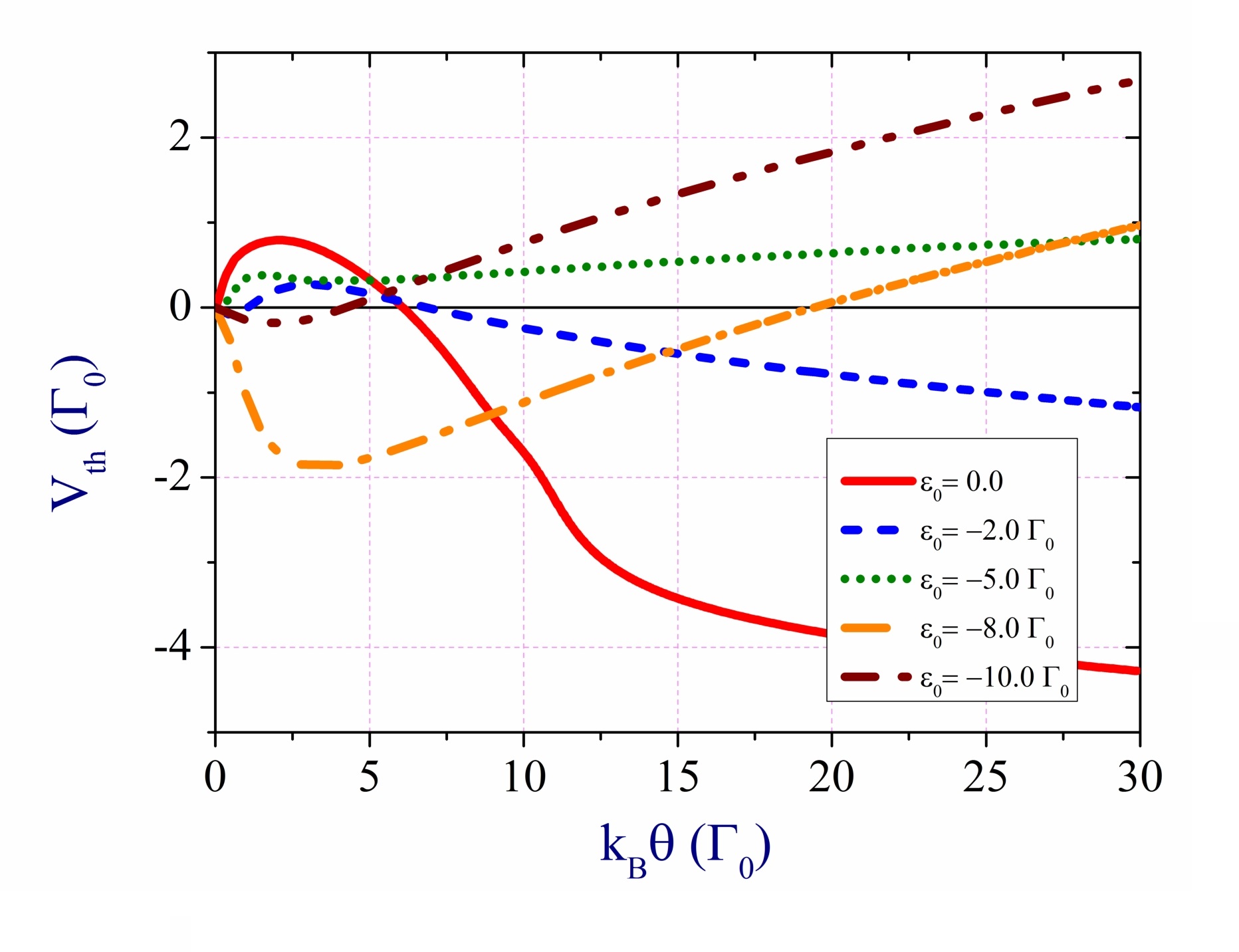}
    \caption{(Color online ) Thermovoltage diagram $V_{\rm th}$ in terms of temperature difference between the right and left electrodes for different energy level of the ${\rm QD}_0$.}
\label{fig5}
\end{figure}

Fig.~\ref{fig5} shows the thermovoltage ($V_{\rm th}$) versus the temperature difference between the right and left electrodes for different  QDs energies. The thermovoltage or Seebeck voltage is obtained by imposing the open-circuit condition $I(V_{\rm th},\theta)=0$, which we extract numerically. 
\par
At very low $\theta$s, thermovoltage is a linear function in terms of temperature that can be positive (for electrons) or negative (for holes) depending on the nature of the charge carriers. While at higher $\theta$, $V_{\rm th}$ shows a nonlinear trend. With raising $\theta$, $V_{\rm th}$'s absolute magnitude increases and gets an extremum. With increasing more the temperature gradient, $V_{\rm th}$ decreases and for a temperature $\theta=\theta_0$, $V_{\rm th}$ becomes zero, and $V_{\rm th}$ changes the sign or $\theta>\theta_0$.
\par
This result is in agreement with experiments on semiconductor quantum dots~\cite{Svensson2013}. The change of the $V_{\rm th}$ sign can be justified as follows: The chemical potential of the electrodes at zero bias voltage is zero. If electrons are the charge carriers, the temperature difference at the electrodes causes the electron to flow from the left electrode (hot) to the right electrode (cold). To reduce this heat current, a negative thermovoltage appears, which grows with increasing $\theta$. Although, with raising $\theta$, the step of the Fermi distribution of the left (hot) electrode is somewhat smoothed such that the hole carries now flow easily to the right electrode. At temperature gradient, $\theta_0$, these flows of electrons and holes become equal and suppress each other leading to a zero heat current. We notice that for the triangle quantum dot setup, considered in this work, at dot energy level $\epsilon_0=-U/2$, thermovoltage is always nonzero and positive, unlike the single quantum dot setup,  which is related to the particle-hole symmetry  broken. 

 \begin{figure}[htb!]
     \includegraphics[width=0.5\textwidth]{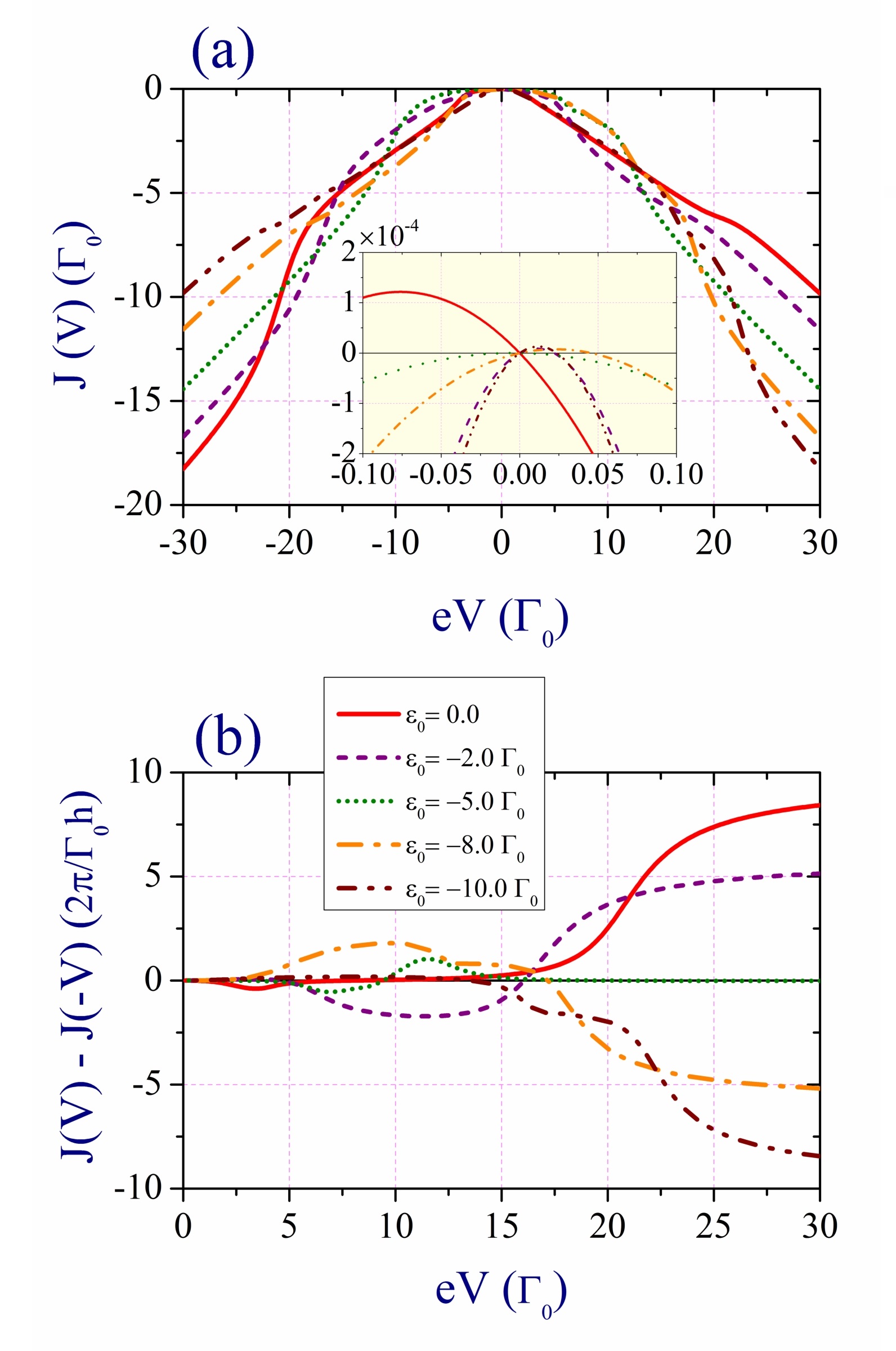}
    \caption{(Color online ) (a) Heat current and (b) the rectification factor as a function of applied voltage in the isothermal case $\theta=0$. Inset: Detail of the dissipated power around zero voltage.}
\label{fig7}
\end{figure}
\par
Materials that have good thermoelectric properties can transform heat into electricity, which is called the Seebeck effect. The Peltier effect is the opposite of the Seebeck effect~\cite{Goldsmid2010}. An electric current that passes through the junction of two substances emits or absorbs heat per unit time at the junction to balance the chemical potential difference between the two substances. The effect of the Peltier, which describes the reversible heat, depends on the direction of the current and, unlike Joule heating, which is irreversible, can be used to cool electrical devices. Recent experimental results indicate that the heat generated at atoms of atomic dimensions shows asymmetric rectification in terms of voltage~\cite{Lee2013,Zotti2014,Naimi2015}. These results are very interesting because while the effects of rectification on the subject of electricity are well known, little is known about how power is lost in mesoscopic conductors under bias voltage.
\par
The linear part of the rectified heat is obtained through the linear response of the Peltier coefficient. Therefore, the power dissipation for positive bias voltage can be greater or less than negative bias voltage, which depends on the placement of resonant energy levels above or below the Fermi energy. Fig.\ref{fig7} shows the heat flow in terms of bias voltage for different amounts of positional energy. As can be seen, for all quantities of dot level positions, the heat flow is an asymmetric function in terms of the voltage sign. In addition, the nonlinear behavioral heat flow  indicates that the effect of higher orders is dominant. It is interesting to note that for all the ${\rm QD}_0$'s energy level positions, the heat flow can be positive or negative (reversible heat) indicating an unusual heat flow for non-zero voltage (whose value depends on the dot level positions)  becomes zero ( see the inset  Fig.\ref{fig7}(a)).  

In Fig. \ref{fig7}(b) we depict the rectification factor $J(V)-J (-V)$ as a function of applied voltage for different dot level positions. For all quantities of dot levels, the sign of the rectifier coefficient changes. For positive values of the rectification factor, the heat dissipation for $V>0$ is higher than the heat dissipation in the $V>0$. At high voltages, $R$ tends to a constant value, because in this range, the Fermi function of the left electrode tends to number one and the Fermi function of the right electrode tends to zero, and the energy flow becomes voltage-independent. 

\section{CONCLUSIONS}\label{sec4}
In this paper, we theoretically investigate the voltage, temperature current and heat dissipation through a triangular quantum dot (QD) attached to the metal electrodes. Calculations were performed using a combination of the non-equilibrium Green function and the equation of motion method in the Coulomb blockade regime. The results show that at the triangular quantum dot, the voltage and temperature current behave nonlinear, and their magnitude and sign depend on factors such as the dot energy level and the hopping parameter between QDs. Based on self-consistent calculations, we showed that the heat current is a nonlinear function concerning the applied voltage. Moreover, the heat current responsing the change of the applied voltage and for all quantities of dot energy levels is an asymmetric function. This feature is unlike the single quantum dot connected to metalic electrodes as reported by aothur in Ref.~\cite{Sierra2014}. We hope the results presented in this work, could shed light on determining the performance of nanoscale electronic devices (in which nonlinear effects are important) to control heat dissipations.

\section{Acknowledgement}  
J. V gratefully acknowledge support from Deutsche Forschungsgemeinschaft (DFG) KE-807/22-1.

\bibliography{ref/references}

\begin{thebibliography}{41}%
\makeatletter
\providecommand \@ifxundefined [1]{%
 \@ifx{#1\undefined}
}%
\providecommand \@ifnum [1]{%
 \ifnum #1\expandafter \@firstoftwo
 \else \expandafter \@secondoftwo
 \fi
}%
\providecommand \@ifx [1]{%
 \ifx #1\expandafter \@firstoftwo
 \else \expandafter \@secondoftwo
 \fi
}%
\providecommand \natexlab [1]{#1}%
\providecommand \enquote  [1]{``#1''}%
\providecommand \bibnamefont  [1]{#1}%
\providecommand \bibfnamefont [1]{#1}%
\providecommand \citenamefont [1]{#1}%
\providecommand \href@noop [0]{\@secondoftwo}%
\providecommand \href [0]{\begingroup \@sanitize@url \@href}%
\providecommand \@href[1]{\@@startlink{#1}\@@href}%
\providecommand \@@href[1]{\endgroup#1\@@endlink}%
\providecommand \@sanitize@url [0]{\catcode `\\12\catcode `\$12\catcode
  `\&12\catcode `\#12\catcode `\^12\catcode `\_12\catcode `\%12\relax}%
\providecommand \@@startlink[1]{}%
\providecommand \@@endlink[0]{}%
\providecommand \url  [0]{\begingroup\@sanitize@url \@url }%
\providecommand \@url [1]{\endgroup\@href {#1}{\urlprefix }}%
\providecommand \urlprefix  [0]{URL }%
\providecommand \Eprint [0]{\href }%
\providecommand \doibase [0]{http://dx.doi.org/}%
\providecommand \selectlanguage [0]{\@gobble}%
\providecommand \bibinfo  [0]{\@secondoftwo}%
\providecommand \bibfield  [0]{\@secondoftwo}%
\providecommand \translation [1]{[#1]}%
\providecommand \BibitemOpen [0]{}%
\providecommand \bibitemStop [0]{}%
\providecommand \bibitemNoStop [0]{.\EOS\space}%
\providecommand \EOS [0]{\spacefactor3000\relax}%
\providecommand \BibitemShut  [1]{\csname bibitem#1\endcsname}%
\let\auto@bib@innerbib\@empty
\bibitem [{\citenamefont {Sothmann}\ \emph {et~al.}(2014)\citenamefont
  {Sothmann}, \citenamefont {S\'anchez},\ and\ \citenamefont
  {Jordan}}]{Sothmann2014}%
  \BibitemOpen
  \bibfield  {author} {\bibinfo {author} {\bibfnamefont {Bj{\"o}rn}\
  \bibnamefont {Sothmann}}, \bibinfo {author} {\bibfnamefont {Rafael}\
  \bibnamefont {S\'anchez}}, \ and\ \bibinfo {author} {\bibfnamefont
  {Andrew~N}\ \bibnamefont {Jordan}},\ }\bibfield  {title} {\enquote {\bibinfo
  {title} {Thermoelectric energy harvesting with quantum dots},}\ }\href
  {\doibase 10.1088/0957-4484/26/3/032001} {\bibfield  {journal} {\bibinfo
  {journal} {Nanotechnology}\ }\textbf {\bibinfo {volume} {26}},\ \bibinfo
  {pages} {032001} (\bibinfo {year} {2014})}\BibitemShut {NoStop}%
\bibitem [{\citenamefont {Urban}(2015)}]{Urban2015}%
  \BibitemOpen
  \bibfield  {author} {\bibinfo {author} {\bibfnamefont {Jeffrey~J.}\
  \bibnamefont {Urban}},\ }\bibfield  {title} {\enquote {\bibinfo {title}
  {Prospects for thermoelectricity in quantum dot hybrid arrays},}\ }\href
  {\doibase 10.1038/nnano.2015.289} {\bibfield  {journal} {\bibinfo  {journal}
  {Nature Nanotechnology}\ }\textbf {\bibinfo {volume} {10}},\ \bibinfo {pages}
  {997--1001} (\bibinfo {year} {2015})}\BibitemShut {NoStop}%
\bibitem [{\citenamefont {Erdman}\ \emph {et~al.}(2017)\citenamefont {Erdman},
  \citenamefont {Mazza}, \citenamefont {Bosisio}, \citenamefont {Benenti},
  \citenamefont {Fazio},\ and\ \citenamefont {Taddei}}]{Erdman2017}%
  \BibitemOpen
  \bibfield  {author} {\bibinfo {author} {\bibfnamefont {Paolo~Andrea}\
  \bibnamefont {Erdman}}, \bibinfo {author} {\bibfnamefont {Francesco}\
  \bibnamefont {Mazza}}, \bibinfo {author} {\bibfnamefont {Riccardo}\
  \bibnamefont {Bosisio}}, \bibinfo {author} {\bibfnamefont {Giuliano}\
  \bibnamefont {Benenti}}, \bibinfo {author} {\bibfnamefont {Rosario}\
  \bibnamefont {Fazio}}, \ and\ \bibinfo {author} {\bibfnamefont {Fabio}\
  \bibnamefont {Taddei}},\ }\bibfield  {title} {\enquote {\bibinfo {title}
  {Thermoelectric properties of an interacting quantum dot based heat
  engine},}\ }\href {\doibase 10.1103/PhysRevB.95.245432} {\bibfield  {journal}
  {\bibinfo  {journal} {Phys. Rev. B}\ }\textbf {\bibinfo {volume} {95}},\
  \bibinfo {pages} {245432} (\bibinfo {year} {2017})}\BibitemShut {NoStop}%
\bibitem [{\citenamefont {Talbo}\ \emph {et~al.}(2017)\citenamefont {Talbo},
  \citenamefont {Saint-Martin}, \citenamefont {Retailleau},\ and\ \citenamefont
  {Dollfus}}]{Talbo2017}%
  \BibitemOpen
  \bibfield  {author} {\bibinfo {author} {\bibfnamefont {Vincent}\ \bibnamefont
  {Talbo}}, \bibinfo {author} {\bibfnamefont {J{\'{e}}r{\^{o}}me}\ \bibnamefont
  {Saint-Martin}}, \bibinfo {author} {\bibfnamefont {Sylvie}\ \bibnamefont
  {Retailleau}}, \ and\ \bibinfo {author} {\bibfnamefont {Philippe}\
  \bibnamefont {Dollfus}},\ }\bibfield  {title} {\enquote {\bibinfo {title}
  {Non-linear effects and thermoelectric efficiency of quantum dot-based
  single-electron transistors},}\ }\href {\doibase 10.1038/s41598-017-14009-4}
  {\bibfield  {journal} {\bibinfo  {journal} {Scientific Reports}\ }\textbf
  {\bibinfo {volume} {7}} (\bibinfo {year} {2017}),\
  10.1038/s41598-017-14009-4}\BibitemShut {NoStop}%
\bibitem [{\citenamefont {De}\ and\ \citenamefont
  {Muralidharan}(2018)}]{De2018}%
  \BibitemOpen
  \bibfield  {author} {\bibinfo {author} {\bibfnamefont {Bitan}\ \bibnamefont
  {De}}\ and\ \bibinfo {author} {\bibfnamefont {Bhaskaran}\ \bibnamefont
  {Muralidharan}},\ }\bibfield  {title} {\enquote {\bibinfo {title} {Non-linear
  phonon peltier effect in dissipative quantum dot systems},}\ }\href {\doibase
  10.1038/s41598-018-23402-6} {\bibfield  {journal} {\bibinfo  {journal}
  {Scientific Reports}\ }\textbf {\bibinfo {volume} {8}} (\bibinfo {year}
  {2018}),\ 10.1038/s41598-018-23402-6}\BibitemShut {NoStop}%
\bibitem [{\citenamefont {Menichetti}\ \emph {et~al.}(2018)\citenamefont
  {Menichetti}, \citenamefont {Grosso},\ and\ \citenamefont
  {Parravicini}}]{Menichetti2018}%
  \BibitemOpen
  \bibfield  {author} {\bibinfo {author} {\bibfnamefont {G}~\bibnamefont
  {Menichetti}}, \bibinfo {author} {\bibfnamefont {G}~\bibnamefont {Grosso}}, \
  and\ \bibinfo {author} {\bibfnamefont {G~Pastori}\ \bibnamefont
  {Parravicini}},\ }\bibfield  {title} {\enquote {\bibinfo {title} {Analytic
  treatment of the thermoelectric properties for two coupled quantum dots
  threaded by magnetic fields},}\ }\href {\doibase 10.1088/2399-6528/aac423}
  {\bibfield  {journal} {\bibinfo  {journal} {Journal of Physics
  Communications}\ }\textbf {\bibinfo {volume} {2}},\ \bibinfo {pages} {055026}
  (\bibinfo {year} {2018})}\BibitemShut {NoStop}%
\bibitem [{\citenamefont {Tang}\ \emph {et~al.}(2018)\citenamefont {Tang},
  \citenamefont {Zhang},\ and\ \citenamefont {Wang}}]{Tang2018}%
  \BibitemOpen
  \bibfield  {author} {\bibinfo {author} {\bibfnamefont {Gaomin}\ \bibnamefont
  {Tang}}, \bibinfo {author} {\bibfnamefont {Lei}\ \bibnamefont {Zhang}}, \
  and\ \bibinfo {author} {\bibfnamefont {Jian}\ \bibnamefont {Wang}},\
  }\bibfield  {title} {\enquote {\bibinfo {title} {Thermal rectification in a
  double quantum dots system with a polaron effect},}\ }\href {\doibase
  10.1103/PhysRevB.97.224311} {\bibfield  {journal} {\bibinfo  {journal} {Phys.
  Rev. B}\ }\textbf {\bibinfo {volume} {97}},\ \bibinfo {pages} {224311}
  (\bibinfo {year} {2018})}\BibitemShut {NoStop}%
\bibitem [{\citenamefont {Ludovico}\ and\ \citenamefont
  {Capone}(2018)}]{Ludovico2018}%
  \BibitemOpen
  \bibfield  {author} {\bibinfo {author} {\bibfnamefont {Mar\'{\i}a~Florencia}\
  \bibnamefont {Ludovico}}\ and\ \bibinfo {author} {\bibfnamefont {Massimo}\
  \bibnamefont {Capone}},\ }\bibfield  {title} {\enquote {\bibinfo {title}
  {Enhanced performance of a quantum-dot-based nanomotor due to coulomb
  interactions},}\ }\href {\doibase 10.1103/PhysRevB.98.235409} {\bibfield
  {journal} {\bibinfo  {journal} {Phys. Rev. B}\ }\textbf {\bibinfo {volume}
  {98}},\ \bibinfo {pages} {235409} (\bibinfo {year} {2018})}\BibitemShut
  {NoStop}%
\bibitem [{\citenamefont {Svilans}\ \emph {et~al.}(2018)\citenamefont
  {Svilans}, \citenamefont {Josefsson}, \citenamefont {Burke}, \citenamefont
  {Fahlvik}, \citenamefont {Thelander}, \citenamefont {Linke},\ and\
  \citenamefont {Leijnse}}]{Svilans2018}%
  \BibitemOpen
  \bibfield  {author} {\bibinfo {author} {\bibfnamefont {Artis}\ \bibnamefont
  {Svilans}}, \bibinfo {author} {\bibfnamefont {Martin}\ \bibnamefont
  {Josefsson}}, \bibinfo {author} {\bibfnamefont {Adam~M.}\ \bibnamefont
  {Burke}}, \bibinfo {author} {\bibfnamefont {Sofia}\ \bibnamefont {Fahlvik}},
  \bibinfo {author} {\bibfnamefont {Claes}\ \bibnamefont {Thelander}}, \bibinfo
  {author} {\bibfnamefont {Heiner}\ \bibnamefont {Linke}}, \ and\ \bibinfo
  {author} {\bibfnamefont {Martin}\ \bibnamefont {Leijnse}},\ }\bibfield
  {title} {\enquote {\bibinfo {title} {Thermoelectric characterization of the
  kondo resonance in nanowire quantum dots},}\ }\href {\doibase
  10.1103/PhysRevLett.121.206801} {\bibfield  {journal} {\bibinfo  {journal}
  {Phys. Rev. Lett.}\ }\textbf {\bibinfo {volume} {121}},\ \bibinfo {pages}
  {206801} (\bibinfo {year} {2018})}\BibitemShut {NoStop}%
\bibitem [{\citenamefont {Josefsson}\ \emph {et~al.}(2018)\citenamefont
  {Josefsson}, \citenamefont {Svilans}, \citenamefont {Burke}, \citenamefont
  {Hoffmann}, \citenamefont {Fahlvik}, \citenamefont {Thelander}, \citenamefont
  {Leijnse},\ and\ \citenamefont {Linke}}]{Josefsson2018}%
  \BibitemOpen
  \bibfield  {author} {\bibinfo {author} {\bibfnamefont {Martin}\ \bibnamefont
  {Josefsson}}, \bibinfo {author} {\bibfnamefont {Artis}\ \bibnamefont
  {Svilans}}, \bibinfo {author} {\bibfnamefont {Adam~M.}\ \bibnamefont
  {Burke}}, \bibinfo {author} {\bibfnamefont {Eric~A.}\ \bibnamefont
  {Hoffmann}}, \bibinfo {author} {\bibfnamefont {Sofia}\ \bibnamefont
  {Fahlvik}}, \bibinfo {author} {\bibfnamefont {Claes}\ \bibnamefont
  {Thelander}}, \bibinfo {author} {\bibfnamefont {Martin}\ \bibnamefont
  {Leijnse}}, \ and\ \bibinfo {author} {\bibfnamefont {Heiner}\ \bibnamefont
  {Linke}},\ }\bibfield  {title} {\enquote {\bibinfo {title} {A quantum-dot
  heat engine operating close to the thermodynamic efficiency limits},}\ }\href
  {\doibase 10.1038/s41565-018-0200-5} {\bibfield  {journal} {\bibinfo
  {journal} {Nature Nanotechnology}\ }\textbf {\bibinfo {volume} {13}},\
  \bibinfo {pages} {920--924} (\bibinfo {year} {2018})}\BibitemShut {NoStop}%
\bibitem [{\citenamefont {Josefsson}\ \emph {et~al.}(2019)\citenamefont
  {Josefsson}, \citenamefont {Svilans}, \citenamefont {Linke},\ and\
  \citenamefont {Leijnse}}]{Josefsson2019}%
  \BibitemOpen
  \bibfield  {author} {\bibinfo {author} {\bibfnamefont {Martin}\ \bibnamefont
  {Josefsson}}, \bibinfo {author} {\bibfnamefont {Artis}\ \bibnamefont
  {Svilans}}, \bibinfo {author} {\bibfnamefont {Heiner}\ \bibnamefont {Linke}},
  \ and\ \bibinfo {author} {\bibfnamefont {Martin}\ \bibnamefont {Leijnse}},\
  }\bibfield  {title} {\enquote {\bibinfo {title} {Optimal power and efficiency
  of single quantum dot heat engines: Theory and experiment},}\ }\href
  {\doibase 10.1103/PhysRevB.99.235432} {\bibfield  {journal} {\bibinfo
  {journal} {Phys. Rev. B}\ }\textbf {\bibinfo {volume} {99}},\ \bibinfo
  {pages} {235432} (\bibinfo {year} {2019})}\BibitemShut {NoStop}%
\bibitem [{\citenamefont {Chi}\ \emph {et~al.}(2020)\citenamefont {Chi},
  \citenamefont {Fu}, \citenamefont {Liu}, \citenamefont {Li}, \citenamefont
  {Wang},\ and\ \citenamefont {Zhang}}]{Chi2020}%
  \BibitemOpen
  \bibfield  {author} {\bibinfo {author} {\bibfnamefont {Feng}\ \bibnamefont
  {Chi}}, \bibinfo {author} {\bibfnamefont {Zhen-Guo}\ \bibnamefont {Fu}},
  \bibinfo {author} {\bibfnamefont {Jia}\ \bibnamefont {Liu}}, \bibinfo
  {author} {\bibfnamefont {Ke-Man}\ \bibnamefont {Li}}, \bibinfo {author}
  {\bibfnamefont {Zhigang}\ \bibnamefont {Wang}}, \ and\ \bibinfo {author}
  {\bibfnamefont {Ping}\ \bibnamefont {Zhang}},\ }\bibfield  {title} {\enquote
  {\bibinfo {title} {Thermoelectric effect in a correlated quantum dot
  side-coupled to majorana bound states},}\ }\href {\doibase
  10.1186/s11671-020-03307-y} {\bibfield  {journal} {\bibinfo  {journal}
  {Nanoscale Research Letters}\ }\textbf {\bibinfo {volume} {15}} (\bibinfo
  {year} {2020}),\ 10.1186/s11671-020-03307-y}\BibitemShut {NoStop}%
\bibitem [{\citenamefont {Ribetto}\ \emph {et~al.}(2021)\citenamefont
  {Ribetto}, \citenamefont {Bustos-Mar\'un},\ and\ \citenamefont
  {Calvo}}]{Ribetto2021}%
  \BibitemOpen
  \bibfield  {author} {\bibinfo {author} {\bibfnamefont {Federico~D.}\
  \bibnamefont {Ribetto}}, \bibinfo {author} {\bibfnamefont {Ra\'ul~A.}\
  \bibnamefont {Bustos-Mar\'un}}, \ and\ \bibinfo {author} {\bibfnamefont
  {Hern\'an~L.}\ \bibnamefont {Calvo}},\ }\bibfield  {title} {\enquote
  {\bibinfo {title} {Role of coherence in quantum-dot-based nanomachines within
  the coulomb blockade regime},}\ }\href {\doibase 10.1103/PhysRevB.103.155435}
  {\bibfield  {journal} {\bibinfo  {journal} {Phys. Rev. B}\ }\textbf {\bibinfo
  {volume} {103}},\ \bibinfo {pages} {155435} (\bibinfo {year}
  {2021})}\BibitemShut {NoStop}%
\bibitem [{\citenamefont {Jong}\ \emph {et~al.}(2021)\citenamefont {Jong},
  \citenamefont {Ri},\ and\ \citenamefont {Ri}}]{Jong2021}%
  \BibitemOpen
  \bibfield  {author} {\bibinfo {author} {\bibfnamefont {Kum~Hyok}\
  \bibnamefont {Jong}}, \bibinfo {author} {\bibfnamefont {Song~Mi}\
  \bibnamefont {Ri}}, \ and\ \bibinfo {author} {\bibfnamefont {Chol~Won}\
  \bibnamefont {Ri}},\ }\bibfield  {title} {\enquote {\bibinfo {title}
  {Parametric study for optimal performance of coulomb-coupled quantum dots},}\
  }\href {\doibase 10.1088/1361-648x/ac0f2a} {\bibfield  {journal} {\bibinfo
  {journal} {Journal of Physics: Condensed Matter}\ }\textbf {\bibinfo {volume}
  {33}},\ \bibinfo {pages} {375302} (\bibinfo {year} {2021})}\BibitemShut
  {NoStop}%
\bibitem [{\citenamefont {Banerjee}\ and\ \citenamefont
  {Singha}(2021)}]{Banerjee2021}%
  \BibitemOpen
  \bibfield  {author} {\bibinfo {author} {\bibfnamefont {Sagnik}\ \bibnamefont
  {Banerjee}}\ and\ \bibinfo {author} {\bibfnamefont {Aniket}\ \bibnamefont
  {Singha}},\ }\bibfield  {title} {\enquote {\bibinfo {title} {A non-local
  cryogenic thermometer based on coulomb-coupled systems},}\ }\href {\doibase
  10.1063/5.0032787} {\bibfield  {journal} {\bibinfo  {journal} {Journal of
  Applied Physics}\ }\textbf {\bibinfo {volume} {129}},\ \bibinfo {pages}
  {114901} (\bibinfo {year} {2021})}\BibitemShut {NoStop}%
\bibitem [{\citenamefont {Dorsch}\ \emph {et~al.}(2021)\citenamefont {Dorsch},
  \citenamefont {Svilans}, \citenamefont {Josefsson}, \citenamefont
  {Goldozian}, \citenamefont {Kumar}, \citenamefont {Thelander}, \citenamefont
  {Wacker},\ and\ \citenamefont {Burke}}]{Dorsch2021}%
  \BibitemOpen
  \bibfield  {author} {\bibinfo {author} {\bibfnamefont {Sven}\ \bibnamefont
  {Dorsch}}, \bibinfo {author} {\bibfnamefont {Artis}\ \bibnamefont {Svilans}},
  \bibinfo {author} {\bibfnamefont {Martin}\ \bibnamefont {Josefsson}},
  \bibinfo {author} {\bibfnamefont {Bahareh}\ \bibnamefont {Goldozian}},
  \bibinfo {author} {\bibfnamefont {Mukesh}\ \bibnamefont {Kumar}}, \bibinfo
  {author} {\bibfnamefont {Claes}\ \bibnamefont {Thelander}}, \bibinfo {author}
  {\bibfnamefont {Andreas}\ \bibnamefont {Wacker}}, \ and\ \bibinfo {author}
  {\bibfnamefont {Adam}\ \bibnamefont {Burke}},\ }\bibfield  {title} {\enquote
  {\bibinfo {title} {Heat driven transport in serial double quantum dot
  devices},}\ }\href {\doibase 10.1021/acs.nanolett.0c04017} {\bibfield
  {journal} {\bibinfo  {journal} {Nano Letters}\ }\textbf {\bibinfo {volume}
  {21}},\ \bibinfo {pages} {988--994} (\bibinfo {year} {2021})}\BibitemShut
  {NoStop}%
\bibitem [{\citenamefont {Zimbovskaya}(2022)}]{Zimbovskaya2022}%
  \BibitemOpen
  \bibfield  {author} {\bibinfo {author} {\bibfnamefont {Natalya~A}\
  \bibnamefont {Zimbovskaya}},\ }\bibfield  {title} {\enquote {\bibinfo {title}
  {Large enhancement of thermoelectric effects in multiple quantum dots in a
  serial configuration due to coulomb interactions},}\ }\href {\doibase
  10.1088/1361-648x/ac640c} {\bibfield  {journal} {\bibinfo  {journal} {Journal
  of Physics: Condensed Matter}\ }\textbf {\bibinfo {volume} {34}},\ \bibinfo
  {pages} {255302} (\bibinfo {year} {2022})}\BibitemShut {NoStop}%
\bibitem [{\citenamefont {Sierra}\ and\ \citenamefont
  {S\'anchez}(2014)}]{Sierra2014}%
  \BibitemOpen
  \bibfield  {author} {\bibinfo {author} {\bibfnamefont {Miguel~A.}\
  \bibnamefont {Sierra}}\ and\ \bibinfo {author} {\bibfnamefont {David}\
  \bibnamefont {S\'anchez}},\ }\bibfield  {title} {\enquote {\bibinfo {title}
  {Strongly nonlinear thermovoltage and heat dissipation in interacting quantum
  dots},}\ }\href {\doibase 10.1103/PhysRevB.90.115313} {\bibfield  {journal}
  {\bibinfo  {journal} {Phys. Rev. B}\ }\textbf {\bibinfo {volume} {90}},\
  \bibinfo {pages} {115313} (\bibinfo {year} {2014})}\BibitemShut {NoStop}%
\bibitem [{\citenamefont {Zimbovskaya}(2015)}]{Zimbovskaya2015}%
  \BibitemOpen
  \bibfield  {author} {\bibinfo {author} {\bibfnamefont {Natalya~A.}\
  \bibnamefont {Zimbovskaya}},\ }\bibfield  {title} {\enquote {\bibinfo {title}
  {The effect of coulomb interactions on nonlinear thermovoltage and
  thermocurrent in quantum dots},}\ }\href {\doibase 10.1063/1.4922907}
  {\bibfield  {journal} {\bibinfo  {journal} {The Journal of Chemical Physics}\
  }\textbf {\bibinfo {volume} {142}},\ \bibinfo {pages} {244310} (\bibinfo
  {year} {2015})}\BibitemShut {NoStop}%
\bibitem [{\citenamefont {Sierra}\ and\ \citenamefont
  {S{\'{a}}nchez}(2015)}]{Sierra2015}%
  \BibitemOpen
  \bibfield  {author} {\bibinfo {author} {\bibfnamefont {Miguel~A.}\
  \bibnamefont {Sierra}}\ and\ \bibinfo {author} {\bibfnamefont {David}\
  \bibnamefont {S{\'{a}}nchez}},\ }\bibfield  {title} {\enquote {\bibinfo
  {title} {Nonlinear heat conduction in coulomb-blockaded quantum dots},}\
  }\href {\doibase 10.1016/j.matpr.2015.05.066} {\bibfield  {journal} {\bibinfo
   {journal} {Materials Today: Proceedings}\ }\textbf {\bibinfo {volume} {2}},\
  \bibinfo {pages} {483--490} (\bibinfo {year} {2015})}\BibitemShut {NoStop}%
\bibitem [{\citenamefont {Zimbovskaya}(2016)}]{Zimbovskaya2016}%
  \BibitemOpen
  \bibfield  {author} {\bibinfo {author} {\bibfnamefont {Natalya~A}\
  \bibnamefont {Zimbovskaya}},\ }\bibfield  {title} {\enquote {\bibinfo {title}
  {Nonlinear thermoelectric transport in single-molecule junctions: the effect
  of electron{\textendash}phonon interactions},}\ }\href {\doibase
  10.1088/0953-8984/28/29/295301} {\bibfield  {journal} {\bibinfo  {journal}
  {Journal of Physics: Condensed Matter}\ }\textbf {\bibinfo {volume} {28}},\
  \bibinfo {pages} {295301} (\bibinfo {year} {2016})}\BibitemShut {NoStop}%
\bibitem [{\citenamefont {S{\'{a}}nchez}\ and\ \citenamefont
  {L{\'{o}}pez}(2016)}]{Sanchez2016}%
  \BibitemOpen
  \bibfield  {author} {\bibinfo {author} {\bibfnamefont {David}\ \bibnamefont
  {S{\'{a}}nchez}}\ and\ \bibinfo {author} {\bibfnamefont {Rosa}\ \bibnamefont
  {L{\'{o}}pez}},\ }\bibfield  {title} {\enquote {\bibinfo {title} {Nonlinear
  phenomena in quantum thermoelectrics and heat},}\ }\href {\doibase
  10.1016/j.crhy.2016.08.005} {\bibfield  {journal} {\bibinfo  {journal}
  {Comptes Rendus Physique}\ }\textbf {\bibinfo {volume} {17}},\ \bibinfo
  {pages} {1060--1071} (\bibinfo {year} {2016})}\BibitemShut {NoStop}%
\bibitem [{\citenamefont {Svilans}\ \emph {et~al.}(2016)\citenamefont
  {Svilans}, \citenamefont {Burke}, \citenamefont {Svensson}, \citenamefont
  {Leijnse},\ and\ \citenamefont {Linke}}]{Svilans2016}%
  \BibitemOpen
  \bibfield  {author} {\bibinfo {author} {\bibfnamefont {Artis}\ \bibnamefont
  {Svilans}}, \bibinfo {author} {\bibfnamefont {Adam~M.}\ \bibnamefont
  {Burke}}, \bibinfo {author} {\bibfnamefont {Sofia~Fahlvik}\ \bibnamefont
  {Svensson}}, \bibinfo {author} {\bibfnamefont {Martin}\ \bibnamefont
  {Leijnse}}, \ and\ \bibinfo {author} {\bibfnamefont {Heiner}\ \bibnamefont
  {Linke}},\ }\bibfield  {title} {\enquote {\bibinfo {title} {Nonlinear
  thermoelectric response due to energy-dependent transport properties of a
  quantum dot},}\ }\href {\doibase 10.1016/j.physe.2015.10.007} {\bibfield
  {journal} {\bibinfo  {journal} {Physica E: Low-dimensional Systems and
  Nanostructures}\ }\textbf {\bibinfo {volume} {82}},\ \bibinfo {pages}
  {34--38} (\bibinfo {year} {2016})}\BibitemShut {NoStop}%
\bibitem [{\citenamefont {G\'omes-Silva}\ \emph {et~al.}(2018)\citenamefont
  {G\'omes-Silva}, \citenamefont {Orellana},\ and\ \citenamefont
  {Anda}}]{Orellana2018}%
  \BibitemOpen
  \bibfield  {author} {\bibinfo {author} {\bibfnamefont {G.}~\bibnamefont
  {G\'omes-Silva}}, \bibinfo {author} {\bibfnamefont {P.~A.}\ \bibnamefont
  {Orellana}}, \ and\ \bibinfo {author} {\bibfnamefont {E.~V.}\ \bibnamefont
  {Anda}},\ }\bibfield  {title} {\enquote {\bibinfo {title} {Enhancement of the
  thermoelectric efficiency in a t-shaped quantum dot system in the linear and
  nonlinear regimes},}\ }\href {\doibase 10.1063/1.5019922} {\bibfield
  {journal} {\bibinfo  {journal} {Journal of Applied Physics}\ }\textbf
  {\bibinfo {volume} {123}},\ \bibinfo {pages} {085706} (\bibinfo {year}
  {2018})}\BibitemShut {NoStop}%
\bibitem [{\citenamefont {Sartipi}\ \emph {et~al.}(2018)\citenamefont
  {Sartipi}, \citenamefont {Hayati},\ and\ \citenamefont
  {Vahedi}}]{Sartipi2018}%
  \BibitemOpen
  \bibfield  {author} {\bibinfo {author} {\bibfnamefont {Zahra}\ \bibnamefont
  {Sartipi}}, \bibinfo {author} {\bibfnamefont {Amir}\ \bibnamefont {Hayati}},
  \ and\ \bibinfo {author} {\bibfnamefont {Javad}\ \bibnamefont {Vahedi}},\
  }\bibfield  {title} {\enquote {\bibinfo {title} {Thermoelectric efficiency in
  three-terminal graphene nano-junctions},}\ }\href {\doibase
  10.1063/1.5044660} {\bibfield  {journal} {\bibinfo  {journal} {The Journal of
  Chemical Physics}\ }\textbf {\bibinfo {volume} {149}},\ \bibinfo {pages}
  {114103} (\bibinfo {year} {2018})}\BibitemShut {NoStop}%
\bibitem [{\citenamefont {Sartipi}\ and\ \citenamefont
  {Vahedi}(2018)}]{Vahedi2018}%
  \BibitemOpen
  \bibfield  {author} {\bibinfo {author} {\bibfnamefont {Z.}~\bibnamefont
  {Sartipi}}\ and\ \bibinfo {author} {\bibfnamefont {J.}~\bibnamefont
  {Vahedi}},\ }\bibfield  {title} {\enquote {\bibinfo {title} {Enhancing
  thermoelectric properties through a three-terminal benzene molecule},}\
  }\href {\doibase 10.1063/1.5018345} {\bibfield  {journal} {\bibinfo
  {journal} {The Journal of Chemical Physics}\ }\textbf {\bibinfo {volume}
  {148}},\ \bibinfo {pages} {174302} (\bibinfo {year} {2018})}\BibitemShut
  {NoStop}%
\bibitem [{\citenamefont {Erdman}\ \emph {et~al.}(2019)\citenamefont {Erdman},
  \citenamefont {Peltonen}, \citenamefont {Bhandari}, \citenamefont {Dutta},
  \citenamefont {Courtois}, \citenamefont {Fazio}, \citenamefont {Taddei},\
  and\ \citenamefont {Pekola}}]{Peltonen2019}%
  \BibitemOpen
  \bibfield  {author} {\bibinfo {author} {\bibfnamefont {P.~A.}\ \bibnamefont
  {Erdman}}, \bibinfo {author} {\bibfnamefont {J.~T.}\ \bibnamefont
  {Peltonen}}, \bibinfo {author} {\bibfnamefont {B.}~\bibnamefont {Bhandari}},
  \bibinfo {author} {\bibfnamefont {B.}~\bibnamefont {Dutta}}, \bibinfo
  {author} {\bibfnamefont {H.}~\bibnamefont {Courtois}}, \bibinfo {author}
  {\bibfnamefont {R.}~\bibnamefont {Fazio}}, \bibinfo {author} {\bibfnamefont
  {F.}~\bibnamefont {Taddei}}, \ and\ \bibinfo {author} {\bibfnamefont {J.~P.}\
  \bibnamefont {Pekola}},\ }\bibfield  {title} {\enquote {\bibinfo {title}
  {Nonlinear thermovoltage in a single-electron transistor},}\ }\href {\doibase
  10.1103/PhysRevB.99.165405} {\bibfield  {journal} {\bibinfo  {journal} {Phys.
  Rev. B}\ }\textbf {\bibinfo {volume} {99}},\ \bibinfo {pages} {165405}
  (\bibinfo {year} {2019})}\BibitemShut {NoStop}%
\bibitem [{\citenamefont {Taniguchi}(2020)}]{Taniguchi2020}%
  \BibitemOpen
  \bibfield  {author} {\bibinfo {author} {\bibfnamefont {Nobuhiko}\
  \bibnamefont {Taniguchi}},\ }\bibfield  {title} {\enquote {\bibinfo {title}
  {Quantum control of nonlinear thermoelectricity at the nanoscale},}\ }\href
  {\doibase 10.1103/PhysRevB.101.115404} {\bibfield  {journal} {\bibinfo
  {journal} {Phys. Rev. B}\ }\textbf {\bibinfo {volume} {101}},\ \bibinfo
  {pages} {115404} (\bibinfo {year} {2020})}\BibitemShut {NoStop}%
\bibitem [{\citenamefont {Goldsmid}(2010)}]{Goldsmid2010}%
  \BibitemOpen
  \bibfield  {author} {\bibinfo {author} {\bibfnamefont {H.~Julian}\
  \bibnamefont {Goldsmid}},\ }\href {\doibase 10.1007/978-3-642-00716-3} {\emph
  {\bibinfo {title} {Introduction to Thermoelectricity}}}\ (\bibinfo
  {publisher} {Springer Berlin Heidelberg},\ \bibinfo {year}
  {2010})\BibitemShut {NoStop}%
\bibitem [{\citenamefont {Staring}\ \emph {et~al.}(1993)\citenamefont
  {Staring}, \citenamefont {Molenkamp}, \citenamefont {Alphenaar},
  \citenamefont {van Houten}, \citenamefont {Buyk}, \citenamefont {Mabesoone},
  \citenamefont {Beenakker},\ and\ \citenamefont {Foxon}}]{Staring1993}%
  \BibitemOpen
  \bibfield  {author} {\bibinfo {author} {\bibfnamefont {A.~A.~M}\ \bibnamefont
  {Staring}}, \bibinfo {author} {\bibfnamefont {L.~W}\ \bibnamefont
  {Molenkamp}}, \bibinfo {author} {\bibfnamefont {B.~W}\ \bibnamefont
  {Alphenaar}}, \bibinfo {author} {\bibfnamefont {H.}~\bibnamefont {van
  Houten}}, \bibinfo {author} {\bibfnamefont {O.~J.~A}\ \bibnamefont {Buyk}},
  \bibinfo {author} {\bibfnamefont {M.~A.~A}\ \bibnamefont {Mabesoone}},
  \bibinfo {author} {\bibfnamefont {C.~W.~J}\ \bibnamefont {Beenakker}}, \ and\
  \bibinfo {author} {\bibfnamefont {C.~T}\ \bibnamefont {Foxon}},\ }\bibfield
  {title} {\enquote {\bibinfo {title} {Coulomb-blockade oscillations in the
  thermopower of a quantum dot},}\ }\href {\doibase 10.1209/0295-5075/22/1/011}
  {\bibfield  {journal} {\bibinfo  {journal} {Europhysics Letters ({EPL})}\
  }\textbf {\bibinfo {volume} {22}},\ \bibinfo {pages} {57--62} (\bibinfo
  {year} {1993})}\BibitemShut {NoStop}%
\bibitem [{\citenamefont {Svensson}\ \emph {et~al.}(2013)\citenamefont
  {Svensson}, \citenamefont {Hoffmann}, \citenamefont {Nakpathomkun},
  \citenamefont {Wu}, \citenamefont {Xu}, \citenamefont {Nilsson},
  \citenamefont {S{\'{a}}nchez}, \citenamefont {Kashcheyevs},\ and\
  \citenamefont {Linke}}]{Svensson2013}%
  \BibitemOpen
  \bibfield  {author} {\bibinfo {author} {\bibfnamefont {S~Fahlvik}\
  \bibnamefont {Svensson}}, \bibinfo {author} {\bibfnamefont {E~A}\
  \bibnamefont {Hoffmann}}, \bibinfo {author} {\bibfnamefont {N}~\bibnamefont
  {Nakpathomkun}}, \bibinfo {author} {\bibfnamefont {P~M}\ \bibnamefont {Wu}},
  \bibinfo {author} {\bibfnamefont {H~Q}\ \bibnamefont {Xu}}, \bibinfo {author}
  {\bibfnamefont {H~A}\ \bibnamefont {Nilsson}}, \bibinfo {author}
  {\bibfnamefont {D}~\bibnamefont {S{\'{a}}nchez}}, \bibinfo {author}
  {\bibfnamefont {V}~\bibnamefont {Kashcheyevs}}, \ and\ \bibinfo {author}
  {\bibfnamefont {H}~\bibnamefont {Linke}},\ }\bibfield  {title} {\enquote
  {\bibinfo {title} {Nonlinear thermovoltage and thermocurrent in quantum
  dots},}\ }\href {\doibase 10.1088/1367-2630/15/10/105011} {\bibfield
  {journal} {\bibinfo  {journal} {New Journal of Physics}\ }\textbf {\bibinfo
  {volume} {15}},\ \bibinfo {pages} {105011} (\bibinfo {year}
  {2013})}\BibitemShut {NoStop}%
\bibitem [{\citenamefont {Wierzbicki}\ and\ \citenamefont
  {\ifmmode~\acute{S}\else \'{S}\fi{}wirkowicz}(2010)}]{Wierzbicki}%
  \BibitemOpen
  \bibfield  {author} {\bibinfo {author} {\bibfnamefont {M.}~\bibnamefont
  {Wierzbicki}}\ and\ \bibinfo {author} {\bibfnamefont {R.}~\bibnamefont
  {\ifmmode~\acute{S}\else \'{S}\fi{}wirkowicz}},\ }\bibfield  {title}
  {\enquote {\bibinfo {title} {Electric and thermoelectric phenomena in a
  multilevel quantum dot attached to ferromagnetic electrodes},}\ }\href
  {\doibase 10.1103/PhysRevB.82.165334} {\bibfield  {journal} {\bibinfo
  {journal} {Phys. Rev. B}\ }\textbf {\bibinfo {volume} {82}},\ \bibinfo
  {pages} {165334} (\bibinfo {year} {2010})}\BibitemShut {NoStop}%
\bibitem [{\citenamefont {Zianni}(2007)}]{Zianni}%
  \BibitemOpen
  \bibfield  {author} {\bibinfo {author} {\bibfnamefont {X.}~\bibnamefont
  {Zianni}},\ }\bibfield  {title} {\enquote {\bibinfo {title} {Coulomb
  oscillations in the electron thermal conductance of a dot in the linear
  regime},}\ }\href {\doibase 10.1103/PhysRevB.75.045344} {\bibfield  {journal}
  {\bibinfo  {journal} {Phys. Rev. B}\ }\textbf {\bibinfo {volume} {75}},\
  \bibinfo {pages} {045344} (\bibinfo {year} {2007})}\BibitemShut {NoStop}%
\bibitem [{\citenamefont {Zimbovskaya}(2020)}]{Zimbovskaya2}%
  \BibitemOpen
  \bibfield  {author} {\bibinfo {author} {\bibfnamefont {Natalya~A}\
  \bibnamefont {Zimbovskaya}},\ }\bibfield  {title} {\enquote {\bibinfo {title}
  {Charge and heat current rectification by a double-dot system within the
  coulomb blockade regime},}\ }\href {\doibase 10.1088/1361-648x/ab83e9}
  {\bibfield  {journal} {\bibinfo  {journal} {Journal of Physics: Condensed
  Matter}\ }\textbf {\bibinfo {volume} {32}},\ \bibinfo {pages} {325302}
  (\bibinfo {year} {2020})}\BibitemShut {NoStop}%
\bibitem [{\citenamefont {Sierra}\ \emph {et~al.}(2016)\citenamefont {Sierra},
  \citenamefont {Saiz-Bret\'{\i}n}, \citenamefont {Dom\'{\i}nguez-Adame},\ and\
  \citenamefont {S\'anchez}}]{Sierra2}%
  \BibitemOpen
  \bibfield  {author} {\bibinfo {author} {\bibfnamefont {Miguel~A.}\
  \bibnamefont {Sierra}}, \bibinfo {author} {\bibfnamefont {M.}~\bibnamefont
  {Saiz-Bret\'{\i}n}}, \bibinfo {author} {\bibfnamefont {F.}~\bibnamefont
  {Dom\'{\i}nguez-Adame}}, \ and\ \bibinfo {author} {\bibfnamefont {David}\
  \bibnamefont {S\'anchez}},\ }\bibfield  {title} {\enquote {\bibinfo {title}
  {Interactions and thermoelectric effects in a parallel-coupled double quantum
  dot},}\ }\href {\doibase 10.1103/PhysRevB.93.235452} {\bibfield  {journal}
  {\bibinfo  {journal} {Phys. Rev. B}\ }\textbf {\bibinfo {volume} {93}},\
  \bibinfo {pages} {235452} (\bibinfo {year} {2016})}\BibitemShut {NoStop}%
\bibitem [{\citenamefont {Meir}\ and\ \citenamefont
  {Wingreen}(1992)}]{Meir1992}%
  \BibitemOpen
  \bibfield  {author} {\bibinfo {author} {\bibfnamefont {Yigal}\ \bibnamefont
  {Meir}}\ and\ \bibinfo {author} {\bibfnamefont {Ned~S.}\ \bibnamefont
  {Wingreen}},\ }\bibfield  {title} {\enquote {\bibinfo {title} {Landauer
  formula for the current through an interacting electron region},}\ }\href
  {\doibase 10.1103/PhysRevLett.68.2512} {\bibfield  {journal} {\bibinfo
  {journal} {Phys. Rev. Lett.}\ }\textbf {\bibinfo {volume} {68}},\ \bibinfo
  {pages} {2512--2515} (\bibinfo {year} {1992})}\BibitemShut {NoStop}%
\bibitem [{\citenamefont {Meir}\ \emph {et~al.}(1991)\citenamefont {Meir},
  \citenamefont {Wingreen},\ and\ \citenamefont {Lee}}]{Meir1991}%
  \BibitemOpen
  \bibfield  {author} {\bibinfo {author} {\bibfnamefont {Yigal}\ \bibnamefont
  {Meir}}, \bibinfo {author} {\bibfnamefont {Ned~S.}\ \bibnamefont {Wingreen}},
  \ and\ \bibinfo {author} {\bibfnamefont {Patrick~A.}\ \bibnamefont {Lee}},\
  }\bibfield  {title} {\enquote {\bibinfo {title} {Transport through a strongly
  interacting electron system: Theory of periodic conductance oscillations},}\
  }\href {\doibase 10.1103/PhysRevLett.66.3048} {\bibfield  {journal} {\bibinfo
   {journal} {Phys. Rev. Lett.}\ }\textbf {\bibinfo {volume} {66}},\ \bibinfo
  {pages} {3048--3051} (\bibinfo {year} {1991})}\BibitemShut {NoStop}%
\bibitem [{\citenamefont {Beenakker}(1991)}]{Beenakker1991}%
  \BibitemOpen
  \bibfield  {author} {\bibinfo {author} {\bibfnamefont {C.~W.~J.}\
  \bibnamefont {Beenakker}},\ }\bibfield  {title} {\enquote {\bibinfo {title}
  {Theory of coulomb-blockade oscillations in the conductance of a quantum
  dot},}\ }\href {\doibase 10.1103/PhysRevB.44.1646} {\bibfield  {journal}
  {\bibinfo  {journal} {Phys. Rev. B}\ }\textbf {\bibinfo {volume} {44}},\
  \bibinfo {pages} {1646--1656} (\bibinfo {year} {1991})}\BibitemShut {NoStop}%
\bibitem [{\citenamefont {Lee}\ \emph {et~al.}(2013)\citenamefont {Lee},
  \citenamefont {Kim}, \citenamefont {Jeong}, \citenamefont {Zotti},
  \citenamefont {Pauly}, \citenamefont {Cuevas},\ and\ \citenamefont
  {Reddy}}]{Lee2013}%
  \BibitemOpen
  \bibfield  {author} {\bibinfo {author} {\bibfnamefont {Woochul}\ \bibnamefont
  {Lee}}, \bibinfo {author} {\bibfnamefont {Kyeongtae}\ \bibnamefont {Kim}},
  \bibinfo {author} {\bibfnamefont {Wonho}\ \bibnamefont {Jeong}}, \bibinfo
  {author} {\bibfnamefont {Linda~Angela}\ \bibnamefont {Zotti}}, \bibinfo
  {author} {\bibfnamefont {Fabian}\ \bibnamefont {Pauly}}, \bibinfo {author}
  {\bibfnamefont {Juan~Carlos}\ \bibnamefont {Cuevas}}, \ and\ \bibinfo
  {author} {\bibfnamefont {Pramod}\ \bibnamefont {Reddy}},\ }\bibfield  {title}
  {\enquote {\bibinfo {title} {Heat dissipation in atomic-scale junctions},}\
  }\href {\doibase 10.1038/nature12183} {\bibfield  {journal} {\bibinfo
  {journal} {Nature}\ }\textbf {\bibinfo {volume} {498}},\ \bibinfo {pages}
  {209--212} (\bibinfo {year} {2013})}\BibitemShut {NoStop}%
\bibitem [{\citenamefont {Zotti}\ \emph {et~al.}(2014)\citenamefont {Zotti},
  \citenamefont {B{\"u}rkle}, \citenamefont {Pauly}, \citenamefont {Lee},
  \citenamefont {Kim}, \citenamefont {Jeong}, \citenamefont {Asai},
  \citenamefont {Reddy},\ and\ \citenamefont {Cuevas}}]{Zotti2014}%
  \BibitemOpen
  \bibfield  {author} {\bibinfo {author} {\bibfnamefont {L~A}\ \bibnamefont
  {Zotti}}, \bibinfo {author} {\bibfnamefont {M}~\bibnamefont {B{\"u}rkle}},
  \bibinfo {author} {\bibfnamefont {F}~\bibnamefont {Pauly}}, \bibinfo {author}
  {\bibfnamefont {W}~\bibnamefont {Lee}}, \bibinfo {author} {\bibfnamefont
  {K}~\bibnamefont {Kim}}, \bibinfo {author} {\bibfnamefont {W}~\bibnamefont
  {Jeong}}, \bibinfo {author} {\bibfnamefont {Y}~\bibnamefont {Asai}}, \bibinfo
  {author} {\bibfnamefont {P}~\bibnamefont {Reddy}}, \ and\ \bibinfo {author}
  {\bibfnamefont {J~C}\ \bibnamefont {Cuevas}},\ }\bibfield  {title} {\enquote
  {\bibinfo {title} {Heat dissipation and its relation to thermopower in
  single-molecule junctions},}\ }\href {\doibase 10.1088/1367-2630/16/1/015004}
  {\bibfield  {journal} {\bibinfo  {journal} {New Journal of Physics}\ }\textbf
  {\bibinfo {volume} {16}},\ \bibinfo {pages} {015004} (\bibinfo {year}
  {2014})}\BibitemShut {NoStop}%
\bibitem [{\citenamefont {Naimi}\ and\ \citenamefont
  {Vahedi}(2015)}]{Naimi2015}%
  \BibitemOpen
  \bibfield  {author} {\bibinfo {author} {\bibfnamefont {Yaghoob}\ \bibnamefont
  {Naimi}}\ and\ \bibinfo {author} {\bibfnamefont {Javad}\ \bibnamefont
  {Vahedi}},\ }\bibfield  {title} {\enquote {\bibinfo {title} {Heat dissipation
  and its relation to molecular orbital energies in single-molecule
  junctions},}\ }\href {\doibase 10.1002/pssb.201552020} {\bibfield  {journal}
  {\bibinfo  {journal} {physica status solidi (b)}\ }\textbf {\bibinfo {volume}
  {252}},\ \bibinfo {pages} {2714--2722} (\bibinfo {year} {2015})}\BibitemShut
  {NoStop}%
\end{thebibliography}%

\end{document}